\documentclass{article}

\pdfpagewidth=\paperwidth
\pdfpageheight=\paperheight

\usepackage{enumerate}
\usepackage{verbatim}
\usepackage{appendix}
\usepackage{comment}
\usepackage{color}

\usepackage{authblk}
\usepackage{amssymb}
\usepackage{amsmath}
\usepackage{mathtools}
\usepackage{amsthm}
\usepackage{amscd}
\usepackage{amsfonts}
\usepackage{graphicx}%
\usepackage{float}
\usepackage{caption}
\usepackage{subcaption}

\begin{document}

\title{Uncertainty quantification in flow cytometry using a cell sorter}

\author{Amudhan Krishnaswamy-Usha, Gregory A. Cooksey,
Paul Patrone}
\affil{
\textit{National Institute of Standards and Technology}, \\ \textit{100 Bureau Drive, Gaithersburg, MD 20899, USA }}

\date{\today}

\maketitle

\begin{abstract}
In cytometry, it is difficult to disentangle the contributions of population variance and instrument noise towards total measured variation. Fundamentally, this is due to the fact that one cannot measure the same particle multiple times. We propose a simple experiment that uses a cell sorter to distinguish instrument-specific variation. For a population of beads whose intensities are distributed around a single peak, the sorter is used to collect beads whose measured intensities lie below some threshold. This subset of particles is then remeasured. If the variation in the measured values is only due to the sample, the second set of measurements should also lie entirely below our threshold. Any ‘spillover’ is therefore due to instrument specific effects - we demonstrate how the distribution of the post-sort measurements is sufficient to extract an estimate of the cumulative variability induced by the instrument. A distinguishing feature of our work is that we do not make any assumptions about the sources of said noise. We then show how `local affine transformations' let us transfer these estimates to cytometers not equipped with a sorter. We use our analysis to estimate noise for a set of three instruments and two bead types, across a range of sample flow rates. Lastly, we discuss the implications of instrument noise on optimal classification, as well as other applications.
\end{abstract}

\section{Introduction}
Flow cytometry is a valuable tool in biotechnology, biomedical research and clinical diagnostics \cite{adan2017flow,ibrahim2007flow,mattanovich2006applications}. Consequently, 
there is a significant body of prior work which aims to quantify properties such as  reproducibility, resolution and background noise of a cytometer \cite{wood1998fundamental,wood1998evaluating,wang2017standardization,steen1992noise,disalvo2022serial}.
{There are multiple sources of measurement variability present in any flow cytometry measurement, such as shot noise due to photon flux in the photodetector \cite{kettlitz2013statistics} or migration across streamlines due to inertial effects at high flow rates \cite{disalvo2022serial}. The distribution of measured intensities is a convolution of these random variables with the underlying distribution of the population, which arises from variation in the number of fluorophores present in each cell/bead. Past work on separating these effects has largely focused on modeling the physics of each separate noise source \cite{chase1998resolution,wood1998fundamental}. In this article, we consider a simple cell sorter based experiment that yields a source-agnostic method of estimating the total instrument-induced variability present in a measurement.}

 A key objective of this work is to overcome unique issues that make uncertainty quantification in cytometry challenging. Chief among these is the inherent lack of \textit{true} reproducibility - each particle can be interrogated only once per laser region. Cell sorters collect cells/particles whose measured fluorescence intensities fall in a user-specified range. This allows us to partially overcome the challenge of reproducibility (or lack thereof). Even though we still cannot repeat the experiment on the same measurand, the sorter allows us to take replicate measurements of arbitrarily chosen sub-populations. As we will show, just two such aggregate measurements are sufficient to obtain an estimate of the { total variation due to instrument and parameter specific effects. }

 A related issue arises when comparing instruments. It is often impossible to compare raw intensity values obtained on different cytometers, especially as the magnitude of such measurements can vary widely for identical samples. We propose a simple affine translation model to harmonize measurements across different instruments. Combined with the noise estimates from the sorter, this allows one to estimate instrument noise present in cytometers that are not equipped with a sorter. 
 
Our analysis relies on two basic assumptions: that the instrument-induced noise is independent of the measured intensity, and that it is Gaussian. This is a \textit{local} assumption, in the sense that we only require it to hold over the range of intensities obtained from a set of single-peak beads. We believe this approximation is reasonable, as the central limit theorem guarantees that the averaged contributions of a large number of independent noise sources converges towards a normal distribution \cite{cramer1999mathematical}. As we discuss in later sections, the independence of the noise does not hold over the entire range of the instrument. In other words, the variance of this noise distribution is actually a function of the measured intensities, i.e. it is constant only locally (cf. \cite{wood1998evaluating}). In order to characterize the complete noise profile of an instrument, we therefore recommend repeating our experiment with tightly clustered sets of beads of varying brightness. 
 
A brief outline of the paper is as follows: we describe our proposed experiment and the accompanying mathematical analysis in Section \ref{methods}. Section \ref{results} contains the results of applying our analysis to three classes of cytometers and two sets of particles of varying intensities. We conclude with a short summary of possible applications, limitations and extensions of our model in Section \ref{disc}.

\section{Methods and mathematical model}\label{methods}
\subsection{Materials and Methods}
`Bead' samples consisted of nominally 7 micron and 15 micron diameter polystyrene spheres that contain fluorophores that emit green light (e.g., 520 nm) when excited by blue light (e.g., 488 nm). Beads were suspended in 1.5 moles/L sodium chloride (to maintain neutral buoyancy) and 0.1\% by volume Triton X-100 as surfactant. A commercial fluorescence activated cell sorter (FACS) instrument configured with a $488$ nm excitation laser and a $527$ nm $\pm$ $16$ nm emission filter was operated at various sample flow rates (1 $\mu$L/min to 200 $\mu$L/min) with gain setting that placed the sample near the middle of the linear-dynamic range for typical `fluorescein' emission collection (e.g., bandpass wavelengths 505 nm to 545 nm). The integrated sum of intensities collected in the fluorescein emission wavelengths while a bead crosses the excitation laser is herein referred to as FITC-A. A threshold intensity was chosen near the peak of the bead's intensity distribution.  Using this threshold, and setting sorting stringency to `single particle', approximately $500,000$ beads were sorted into a collection vial.  Following sorting, the collection vial was either 1) resampled through the FACS instrument (``Instrument 1''), 2) measured on a commercial analytical cytometer (``Instrument 2"), or 3) measured on a custom microfluidic cytometer (``Instrument 3"). Instrument 2 was configured with $488$ nm excitation laser and $525$ nm $\pm$ $20$ nm emission filter and utilizes an avalanche photodiode for detection.  Sample flow rates were varied between 10 $\mu$L/min and 90 $\mu$L/min.  Instrument 3 is described in \cite{disalvo2022serial}, and was configured with a $488$ nm excitation laser and $520$ nm $\pm 20$  nm emission filter and utilizes a photomultiplier tube for detection.  Sample flow rates were run at approximately $1$ $\mu$L/min.

\subsection{The cell sorter experiment}

 Assume that the population consists of a set of beads containing a nominal fluorophore concentration, whose fluorescence intensity is normally distributed around a mean value. The distribution of the measured intensities (the integrated area of the voltage pulse from the photodetector) has two broad contributing factors i) variation in the number of flurophores per bead, and ii) variability due to all other sources. Throughout what follows, we will call the distribution induced entirely due to population variability the \textit{signal} and the distribution induced by all other instrument related sources the \textit{noise}. Mathematically, if $Y$ is the random variable denoting the measured fluorescence intensity of a particle, and $X$ is the random variable corresponding to the `true' fluorescence intensity of the particle, we may write $Y=X+\epsilon$, where $\epsilon$ is the random variable corresponding to the cumulative noise from all sources.  Note that these sources are not limited to effects such as shot noise and amplification, but also encompasses phenomena such as how flow instability affects signal intensity.

\begin{figure}
     
\begin{subfigure}[b]{0.45\linewidth}
    \includegraphics[width=\linewidth]{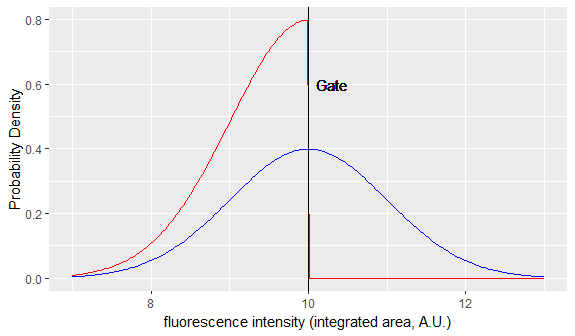}
    \caption{}
    \end{subfigure}
\begin{subfigure}[b]{0.45\linewidth}
    \includegraphics[width=\linewidth]{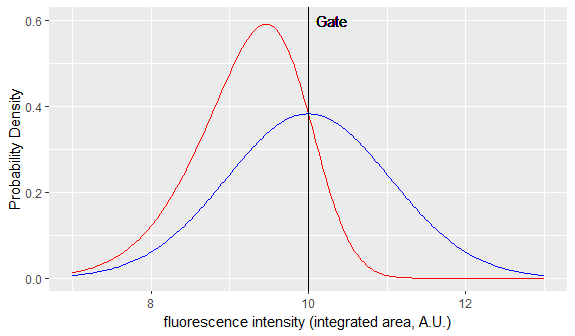}
    \caption{}
    \end{subfigure}
\begin{subfigure}[b]{\linewidth}
    \includegraphics[width=\linewidth]{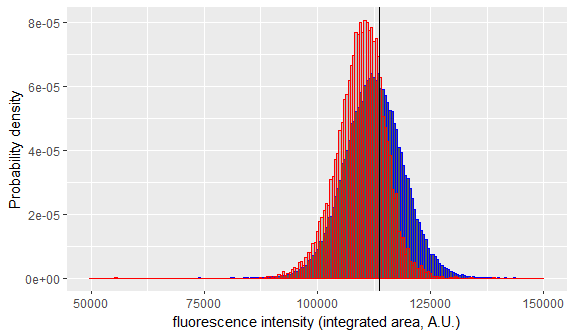}
       
\end{subfigure}
\caption{ Idealized pre-sort (a) and post-sort (b) probability density functions of measured fluorescence intensities, compared with the results of an experiment using a set of 7 micron beads (c). The sharp discontinuity in the density function in (a) is due to the absence of instrument noise, while the spillover above the gate in (b) is due to instrument noise equal to 30\% of the population variance. The gate in both cases is chosen to be equal to the population mean.  The \textit{gate} is the vertical line, and the gating region is all intensities lying below the gate. }
\label{fig:15umpdf}
\end{figure}

A sorter allows us to selectively collect beads with measured intensities in a specified \textit{gating region}. For what follows, we assume that the \textit{purity} of the sorter is close to $100\%$ \cite{bdtech}. In other words, the probability of collecting a bead whose measured intensity lies outside the gating region is low. Note that is not a statement about the \textit{yield} - i.e, we do not require that every bead whose measured intensity lies in the gating region is collected. We assume that the probability of capturing a bead is independent of its measured value (constrained on the value being in the gating region). This implies that the collected set of beads is a sub-sample of the set of all beads whose intensities lie in our chosen region. We choose a \textit{gate} $\gamma$ from the middle of the measured distribution and collect beads whose intensities lie below the gate.  Let $Z$ denote the random variable corresponding to the distribution one obtains by re-measuring this set of beads. In the absence of all noise ($\epsilon=0$), the distribution of $Z$ will exhibit a sharp discontinuity, with no values above $\gamma$. In reality, we observe some `spillover' of values above $\gamma$, which arises due to the noise present in the first and second measurements (see Figure \ref{fig:15umpdf} for an illustration). 

 Our goal is to use the information contained in the distribution of post-sort intensities to estimate the instrument noise parameter $\epsilon$. To do so, we first derive an explicit form for the post-sort cumulative distribution function (CDF), which we denote by $K(x)$. To simplify calculations, we make a couple of assumptions. First, we assume that the distributions of $X$ and $\epsilon$ are both Gaussian. This is a reasonable assumption for $X$ given our choice of bead formulation. The justification for $\epsilon$ is that the contributions of sufficiently many independent sources of noise ought to converge in distribution to a Gaussian. At this stage, we also assume that $\epsilon$ has zero mean (we will revisit this assumption later, when dealing with cross-instrument harmonization).  
 Our second assumption is that the noise is locally independent of the population variable $X$. Since the variance of the noise is likely to depend on the true fluorescence intensity, this is only a first approximation. This is justified in our case since the fluorescence intensities of our chosen beads occupy a narrow band of values. 

 Consider a thought experiment where we possess the ability to track a single bead $\omega$ through the course of our experiment. Further, assume that the bead $\omega$ gets selected by the sorter and re-measured. Let $Y_1=Y_1(\omega)$ and $Y_2=Y_2(\omega)$ denote the first and second measurements of $\omega$. Since the underlying parameters and noise sources remain the same, we assume that $Y_1= X + \epsilon_1, Y_2= X+\epsilon_2$ where $\epsilon_1,\epsilon_2$ denote the independent and identically distributed noise added in the first and second measurement respectively. To obtain the post-sorting cumulative distribution function of intensity values, we compute the probability that $Y_2$ lies below some value $x$ given that $Y_1$ lies below the gate $\gamma$. In other words, if we let $P(A)$ denote the probability of a set of events $A$,
\[
K(x) = P(Y_2 < x | Y_1 < \gamma).
\]
By a direct application of Bayes' rule of conditional probability \cite{cramer1999mathematical}, one obtains
\begin{equation} K(x) =  P(Y_2 <x | Y_1 < \gamma) = \frac{P( X + \epsilon_2 <x, X + \epsilon_1<\gamma)}{P(X+\epsilon_1<\gamma)}.
\end{equation}
The numerator represents the probability that two independent measurements of a particular bead on the same instrument yield values less than $x$ and $\gamma$ respectively. Since $X+\epsilon_2$ and $X+\epsilon_1$ are both normal random variables, the numerator and denominator can be expressed in terms of more familiar functions. 
In particular, let $\phi$ denote the probability density function (PDF) of the standard normal distribution and let $\Phi$ denote its cumulative density function (CDF).\footnote{If $\phi$ and $\Phi$ are the PDF and CDF of a standard normal distribution, \[\phi(x) = \frac{1}{\sqrt{2 \pi}} e^{-\frac{x^2}{2}}, \;\; \Phi(x) = \int_{-\infty}^x \phi(t) dt\]} Let $\mu,\lambda^2$ be the mean and variance of the random variable $X$, and let $\sigma^2$ be the variance of $\epsilon_1$ and $\epsilon_2$. Then the pre-sort measurement $Y_1=X+\epsilon_1$ is a normal random variable with mean $\mu$ and variance $\nu^2=\lambda^2+\sigma^2$. Expressed in terms of the standard normal CDF $\Phi$, we have \[P(X+\epsilon_1<\gamma)=\Phi\left(\frac{\gamma-\mu}{\nu}\right).\]
By scaling our random variables, we may write
\[
P(X+\epsilon_2<x,X+\epsilon_1<\gamma)= P\left( \frac{X+\epsilon_2-\mu}{\nu}< \frac{x-\mu}{\nu}, \frac{X+\epsilon_1 - \mu}{\nu}<\frac{\gamma-\nu}{\nu}\right).\]
Observe that two consecutive measurements are not independent random variables, as these measurements come from overlapping sets of beads. Using the linearity of expectations, the independence of $X,\epsilon_1,\epsilon_2$, and the fact that $\epsilon_1,\epsilon_2$ have mean zero to compute the covariance between these random variables, we obtain
\begin{eqnarray} \textrm{Cov}\left(\frac{X+\epsilon_2-\mu}{\nu}, \frac{X+\epsilon_1-\mu}{\nu}\right) &= &\mathbb{E}\left(\frac{X-\mu+\epsilon_1}{\nu} \frac{X-\mu+\epsilon_2}{\nu} \right) \nonumber \\ &=& \mathbb{E}\left(\frac{(X-\mu)^2+(\epsilon_1+\epsilon_2)(X-\mu)+\epsilon_1\epsilon_2}{\nu^2}  \right)\nonumber\\
&=&\frac{\textrm{Var}(X)}{\nu^2}=\frac{\lambda^2}{\nu^2}.\end{eqnarray}

In other words, the probability $P(X+\epsilon_1 < \gamma, X+\epsilon_2 < x)$ is precisely the probability that two normal random variables with mean $0$, variance $1$ and covariance $\rho=\frac{\lambda^2}{\nu^2}$ have values less than $\frac{\gamma-\mu}{\nu}$ and $\frac{x-\mu}{\nu}$ respectively.
Analytically, if we let $\Phi_2(x,y,\rho)$ denote the joint cumulative distribution at $(x,y)$ of two standard normal random variables with covariance $\rho$, we have \[ \Phi_2(x,y,\rho)= \frac{1}{2\pi \sqrt{1-\rho^2}}\int_{-\infty}^x \int_{-\infty}^y e^{-\frac{t^2-2\rho s t +s^2}{2(1-\rho^2)}} ds \,dt. \]
Therefore, 
\[P(X+\epsilon_2<x,X+\epsilon_1<\gamma)=\Phi_2\left( \frac{\gamma-\mu}{\nu}, \frac{x-\mu}{\nu}, \frac{\lambda^2}{\nu^2}\right),
\]
which gives us the following expression for the post-sort cumulative distribution: 
\begin{equation}\label{eqn:Kbivar}
K_{\gamma,\mu,\nu,\lambda}(x) = K(x)= \Phi_2\left( \frac{\gamma-\mu}{\nu}, \frac{x-\mu}{\nu}, \frac{\lambda^2}{\nu^2}\right) \times \frac{1}{\Phi\left(\frac{\gamma-\mu}{\nu}\right)}.
\end{equation}

We note as an aside that distributions of the form in \eqref{eqn:Kbivar} arise in the statistical theory of hidden truncation models \cite{arnold2000hidden}, although to our knowledge they have not been utilized in noise estimation problems.   
We draw attention to the fact that the post-sort distribution at $x$ depends on only three parameters: the $Z$-score (i.e. distance from the mean as measured in terms of standard deviations) of the gate $\gamma$, the $Z$-score of $x$, and the ratio of the signal and the measured variances. We are interested in using this information to obtain the true population-level variance $\lambda^2$ (or equivalently, the noise variance $\sigma^2=\nu^2-\lambda^2$, since we may estimate $\nu^2$ from the data). Note that the post-sort cumulative distribution function $K$ can be estimated by the empirical CDF $\widehat{K}$. $\widehat{K}$ is defined by 
\begin{equation}
    \widehat{K}(x) = \frac{1}{N} |\{f_j: f_j <x\}|,
\end{equation}
where $f_1,\ldots,f_N$ are the measured post-sort fluorescence intensities, and $|A|$ denotes the cardinality of a set $A$.
Since $\gamma$ is a parameter we choose, and $\mu,\nu$ and $K(x)$ can all be estimated from the data, this allows us to construct an estimator $\widehat{\lambda^2}$ for $\lambda^2$ by optimising for the difference between the predicted and the empirical CDF \footnote{We opt for this instead of a more conventional maximum likelihood estimator as it is much easier to numerically compute the post-sort CDF as compared to the PDF. Additionally, numerical evidence suggests that under the conditions of our model, the estimator we use converges faster than other conventional CDF-based estimators, such as the maximum spacing estimator.}:
\begin{equation}\label{eqn:SDEstimator}
\widehat{\lambda^2}: = \mathop{\textrm{argmin}}_{s \in (0,\nu)} \sum_{i=1}^n \left(K_{\gamma,\hat{\mu},\hat{\nu},s}(x_i)- \widehat{K}(x_i) \right)^2.
\end{equation}
Here $\hat{\mu},\hat{\nu}$ are the sample mean and standard deviation of the pre-sort measurement, $x_1\ldots x_n$ are a grid of points (our analysis uses $10$ points from $-1$ to $+1$ standard deviations of the mean), and $\widehat{K}(x_i)$ is the empirical estimator of the cumulative post-sort distribution. 
\subsection{Additional considerations}
\begin{figure}
\begin{subfigure}[b]{\linewidth}
    \includegraphics[width=\linewidth]{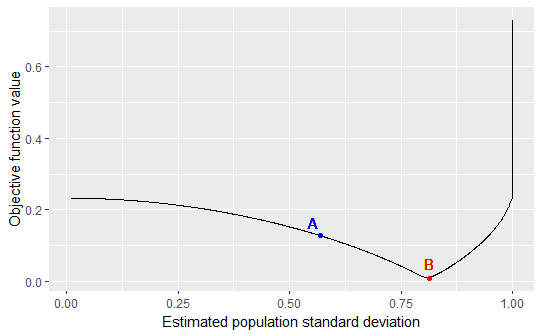}
\end{subfigure}
  \begin{subfigure}[b]{\linewidth}
      \includegraphics[width=\linewidth]{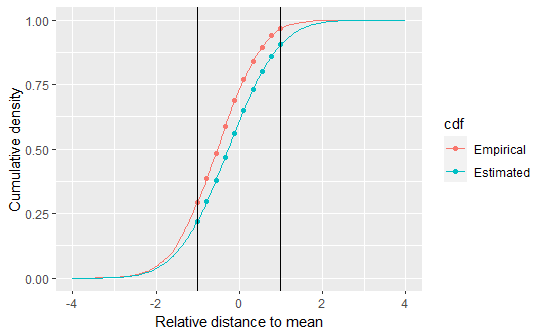}
  \end{subfigure}  
  \caption{(Top) Objective function value versus the estimated population standard deviation. Point B on the graph marks the minimum of the objective function. (Bottom) Graph of the empirical vs estimated CDF corresponding to an estimated population standard deviation of 0.56. The objective function minimizes the distance between the grid of points shown between the black vertical lines. The estimated CDF for point B differs from the empirical CDF by less than $1\%$ and is not shown.  Plots shown were generated using synthetic data, with Gaussian signal and noise of mean $0$ and variance $0.8$ and $0.6$ respectively, along with a gating threshold of $0$.} 
    \label{fig:objfnplot}
\end{figure}

Experimental evidence (see Figure \ref{fig:15hist}) suggests that post-sort measurements exhibit a slight downward shift in intensity unaccounted for by our model, which we speculate is due to bleaching of the fluorophores \cite{van1992photo}. As a first approximation, we assume that the bleaching affects the bead population uniformly. We therefore modify our estimator \eqref{eqn:SDEstimator} to correct for this, by replacing the empirical estimator of the cdf $\widehat{K}(x_i)$ by a translated version $\widehat{K}(x_i+\delta)$, and optimizing for both the shift $\delta$ and the population-level variance $\lambda^2$. 

\begin{equation}\label{eqn:SDEstimator_offset}
(\widehat{\lambda^2},\delta) = \mathop{\textrm{argmin}}_{(s,t)\in (0,\nu)\times \mathbb{R}} \sum_{i=1}^n \left(K_{\gamma,\hat{\mu},\hat{\nu},s}(x_i)- \widehat{K}(x_i+t) \right)^2.
\end{equation}
Figure \ref{fig:objfnplot} shows the typical shape of this objective function on one axis, along with some values of the empirical and estimated CDFs. 

We omit the derivation for the sake of conciseness, but we note here that the probability density of the post-sort distribution has the form
\begin{equation}\label{eqn:Nk}
    k(x) = C^{-1} \int_\mathbb{R} \frac{1}{\sigma}\phi\left(\frac{x-r}{\sigma}\right) \frac{1}{\lambda} \phi\left(\frac{r-\mu}{\lambda}\right) \Phi\left(\frac{\gamma-r}{\sigma}\right) dr =(h \star \alpha)(x),
\end{equation}
where 
\begin{equation}\label{eqn:Nalpha}
    h(x) =\frac{1}{\sigma}\phi\left(\frac{x}{\sigma}\right),  \, \alpha(r)=C^{-1}\frac{1}{\lambda} \phi\left(\frac{r-\mu}{\lambda}\right) \Phi\left(\frac{\gamma-r}{\sigma}\right) , \; \; C=\Phi\left(\frac{\gamma-\mu}{\nu}\right).
\end{equation}

From equation \eqref{eqn:Nk}, we see that the post-sort density is equal to that of a random variable $W+\epsilon_2$, where $W$ has density $\alpha$ and is independent of $\epsilon_2$. This reflects the fact that the post-sort measurements contain two potential sources of variation - the $\epsilon_2$ term corresponds to measurement uncertainty, while the other is a misclassification error due to noise from the first measurement, which is baked into equation \eqref{eqn:Nalpha}.

\subsection{Cross-instrument comparisons}

The analysis so far has assumed that the pre and post-sort measurements are both done on the FACS. However, we could potentially re-measure the pre and post-sort populations on different instruments. Given the wide variability of instrument configurations and settings, it is often the case that cytometers measure fluorescence on different scales. In order to meaningfully compare results from different cytometers, we make the modeling assumption that the true signal in both instruments are related via an affine transformation. As before, denote the measured value in instrument 1 (the sorter) by $X+\epsilon_1$, with $X$ the signal and $\epsilon_1$ the noise. We now assume that the measured value in instrument 2 is of the form $AX+B + \epsilon_2$, where $A,B$ are constants, $B+\epsilon_2$ is the noise inherent to instrument 2 and $\epsilon_2$ is a $\mathcal{N}(0,\sigma_2^2)$ random variable.\footnote{For the sorter, we safely made the assumption that the noise had mean $0$, as arbitrary constant shifts in the measured values are irrelevant for practical purposes. We incorporate the term $B$ when dealing with different instruments to account for the fact that the \textit{relative} shift in each instrument might be different.}
(see \cite{patrone2020affine} for similar approaches to cross-instrument comparisons in a different context). The derivation in the previous section carries over mutatis mutandis to give us the post-sort distribution on instrument 2.
\begin{equation}\label{eqn:KII}
K_{2}(x) = \Phi_2\left( \frac{\gamma-\mu}{\nu_1}, \frac{x-B-A\mu}{\nu_2}, \frac{A\lambda^2}{\nu_1 \nu_2}\right) \times \frac{1}{\Phi\left(\frac{\gamma-\mu}{\nu_1}\right)}.
\end{equation}
Here $\gamma$ is the gating value, $\mu$ is the mean of the pre-sort population measured on instrument 1, $\nu_1,\nu_2$ are the standard deviations of the pre-sort measurements on instrument 1 and 2 respectively, and $\lambda^2$ is the true population variance, estimated using the sorter.  

Our goal here is to estimate $A$ and $B$ using the mean and variance of the pre and post-sort distributions. Note that equation \eqref{eqn:KII} shows that the affine relationship between the pre-sort population distributions carries over to the post-sort distributions as well. The post-sort distribution on instrument 1 is $W+\epsilon_1$ ($W$ is a random variable with density given by equation \eqref{eqn:Nalpha})  while the post-sort distribution on instrument 2 is $AW+B+\epsilon_2$. As a result, we have a simple way of estimating the coefficients $A$ and $B$:

\begin{equation}\label{eqn:Aest}
A = \frac{\mu_{pre,2} - \mu_{post,2}}{\mu_{pre,1}-\mu_{post,2}},
\end{equation}
\begin{equation}\label{eqn:Best}
B = \mu_{pre,2} - A\mu_{pre,1},
\end{equation}
where $\mu_{pre,1},\mu_{post,1}$, $\mu_{pre,2},\mu_{post,2}$ denote the sample mean of the pre and post sort measurements on instruments 1 and 2 respectively. 

With $A$ and $B$ in hand, we can estimate $\sigma_2^2$, the variance of the noise associated with instrument 2 once we have an estimate of the noise from instrument 1. 

\begin{equation}\label{eqn:sigma2}
\sigma_2^2 = \nu_2^2 - A^2 (\nu_1^2 - \sigma_1^2),
\end{equation}
where $\nu_1^2,\nu_2^2$ are the sample variances of the pre-sort measurements on instrument 1 and 2, and $\sigma_1^2$ is the estimated noise variance on instrument 1. 

This provides us a method of comparing the noise present in different cytometers. Equation \eqref{eqn:KII} also provides a way of validating our assumptions and analysis, by comparing our predictions with the empirical post-sort CDF on instrument 2.
Alternatively, since we have three unknown parameters, we might estimate $A,B$ and $\lambda$ by constructing a 3D objective function which minimizes the difference in the empirical and theoretical CDFs of the post-sort distribution on instrument 2. As we will show in later sections, our approach also results in concurrence between the CDFs, and is likely equivalent to the method outlined above.

\section{Results}\label{results}

\subsection{Numerical validation and estimator properties}
\begin{figure}[H]
\begin{subfigure}[b]{0.5\linewidth}
    \includegraphics[width=\linewidth]{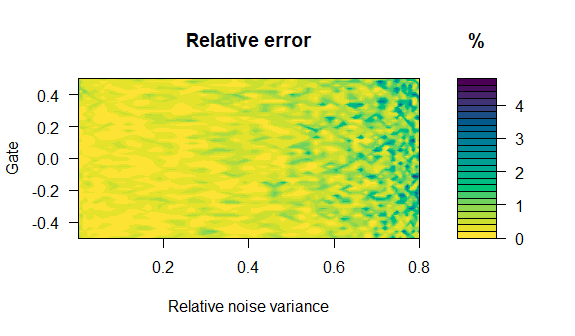}
    \caption{}
    \end{subfigure}
\begin{subfigure}[b]{0.5\linewidth}
    \includegraphics[width=\linewidth]{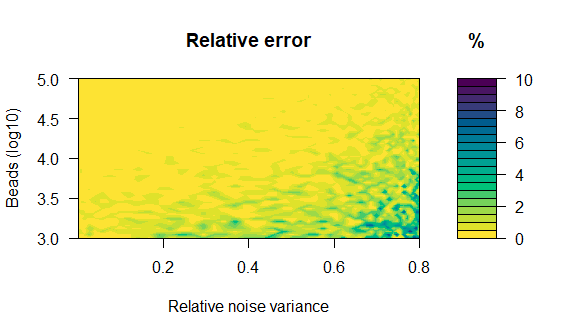}
    \caption{}
 \end{subfigure}
  \caption{Contour plots of the relative error of the estimated signal (i.e. population) variance for synthetic data. Figure (a) depicts the results for a pre-sort bead population of $10,000$, as a function of the gate position (in standard deviations from the mean) and the relative noise variance ($\frac{\sigma^2}{\nu^2}$, in the notation of the previous section). Figure (b) shows the dependence of the relative noise variance on the count of the pre-sort bead population, when the gate is equal to the mean.} 
    \label{fig:synth}
\end{figure}

The estimator in equation \eqref{eqn:SDEstimator} is asymptotically consistent, as it can be expressed as a continuous function of other consistent or unbiased estimators \cite{cramer1999mathematical}. In other words, given any interval around the true population variance $\lambda$, the probability that the estimated variance $\widehat{\lambda}$ lies in that interval goes to $1$ as the number of data points goes to infinity. Numerical simulations (Figure \ref{fig:synth}) also appear to indicate that the relative error (i.e. $\frac{|{\mathbb{E}(\widehat{\lambda})-\lambda}|}{\lambda}$, where $\mathbb{E}$ denotes the expected value over multiple trials) of this estimator is below $5\%$, provided the pre-sort population size is at least of the order of $10^4$, and the instrument contribution to the measured variance is low. 

 \subsection{Noise estimates for a sorter}

\begin{figure}
    \centering    \includegraphics[width=\linewidth]{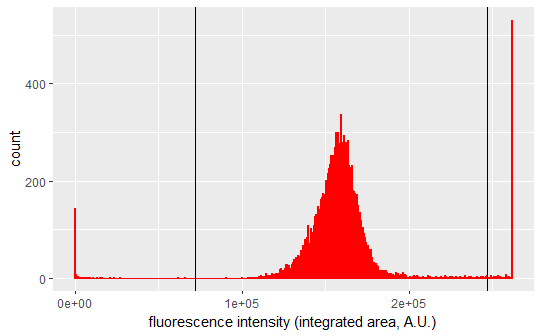}
    \caption{Histogram of the untruncated pre-sort measurements of a set of 15 micron beads. The vertical lines depict the cut-offs used to discard spurious events and multiplets, and correspond to $\pm 3$ standard deviations of the mean.}
    \label{fig:15trunc}
\end{figure}

\begin{figure}
    \centering
\includegraphics[width=\linewidth]{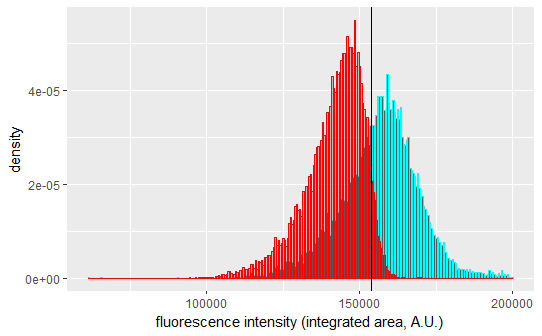}
    \caption{Pre (cyan) and post (red) sort measurements for a set of 15 micron beads. The black vertical line is the gate that was used for sorting. The red curve exhibits an offset relative to what the theory would predict based on sorting and instrument noise alone. We hypothesize that this offset is due to photobleaching of the fluorophores. Note that any sorting errors should result in a rightward shift of the distribution, due to the inclusion of beads with measured intensity above the gate.}
    \label{fig:15hist}
\end{figure}

\begin{figure}
\begin{subfigure}[b]{0.5\linewidth}
\includegraphics[width=\linewidth]{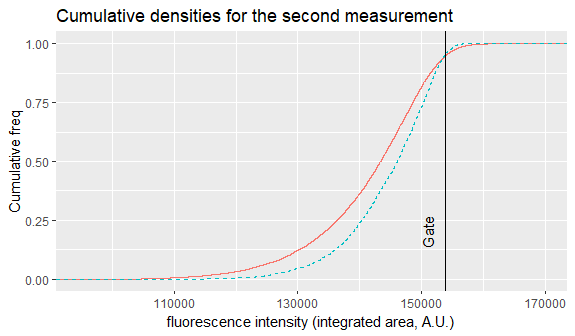}
\end{subfigure}
\begin{subfigure}[b]{0.5\linewidth}
    \includegraphics[width=\linewidth]{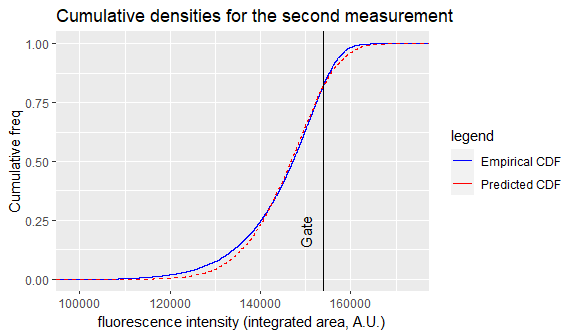}
  \end{subfigure}  
     \caption{Predicted vs emprical CDF of the post sort distribution for the $15$ micron beads from Figure \ref{fig:15hist}. (a) shows the results of applying the estimator  \eqref{eqn:SDEstimator}, which does not correct for the observed offset. (b) shows the results of using the objective function \eqref{eqn:SDEstimator_offset}, which results in a maximum difference of $<2\%$ between the predicted and observed CDFs.}	
    \label{fig:cdf_15um}
\end{figure}

Two sets of beads ($7\mu\text{m}$ and $15 \mu\text{m}$ diameter beads labeled with fluorophores excited by 488 nm wavelength lasers) were subjected to the sorting experiment described above. Informed by our numerical simulations, we chose the sorting gate to be within $1$ standard deviation of the population mean (Table \ref{table:sorterraw}). To eliminate spurious outliers from our data (due to multiplets or `empty' events), we truncate the observed pre and post-sort distributions to $\pm 3$ standard deviations of the mean of the pre-sort measurements (Figure \ref{fig:15trunc}).
Figure \ref{fig:15hist} depicts the pre and post-sort distributions of the same set of 15 micron beads, post truncation. Figure \ref{fig:cdf_15um} shows the results of applying our estimators \eqref{eqn:SDEstimator} and \eqref{eqn:SDEstimator_offset} to the data shown in Figure \ref{fig:15hist}.

\begin{table}[h]
\centering

\caption{Estimated population and instrument induced standard deviations for different bead types measured on the sorter. Flow rate is the estimated sample flow rate, in $\mu$L/min. $\mu$ and $\nu$ are the measured pre-sort mean and standard deviation (fluorescence intensity, integrated area, a.u), $\sigma$ is the estimated standard deviation attributable to instrument noise. $\mu,\gamma,\nu,\sigma$ all represent fluorescence intensity (integrated area, A.U.) } 
\begin{tabular}{ c c| c c c c }
 Bead size & Flow rate & Pop. mean ($\mu$) & Gate ($\gamma$) & Total SD($\nu)$ & Noise. SD($\sigma$) \\
\hline
 $7 \mu\text{m}$ & 22 $\mu$L/min &112373 & 113637 & 5953 & 2736\\ 
 $15 \mu\text{m}$ & 29.5$\mu$L/min &156591 & 153891 & 12724 & 2960  
\end{tabular}
\label{table:sorterraw}
\end{table}

It is also useful to consider other metrics of uncertainty, such as the \textit{relative noise variance} $\frac{\sigma^2}{\nu^2}$, the ratio of the noise variance to the measured variance. This is a measure of how much the spread of the observed intensities arises from the instrument as opposed to variation in the beads. If two sets of beads with approximately similar intensity ranges are measured on a cytometer under identical conditions, the resulting relative noise variances serve as a proxy for bead quality, as lower values imply that the spread in observed intensities is almost entirely due to external noise. This likely indicates that one set of beads has a tighter distribution of bead shapes and fluorophore concentrations.

Another quantity of interest is the \textit{relative error}, the ratio of the  standard deviation of the noise to the empirical mean ($\textrm{Rel. Error}=\sigma/\mu$). This is a measure of the averaged relative error across individual measurements. As seen in Table \ref{table:sorter}, the relative error is fairly small for both bead sizes, approximately between $1\%$ to $ 2\%$. On the other hand, the relative noise variance is much lower for the larger (and hence brighter) beads, which indicates that the spread seen in the intensities for these beads is almost entirely due to variation in the beads themselves. Table \ref{table:sorter} summarises the derived quantities described above for different bead sizes on the sorter.

\begin{table}[h]
\centering
\caption{Estimated metrics of uncertainty for different bead sizes on the sorter. $\mu$ and $\nu^2$ are the measured mean and variance of the pre-sort distribution, $\sigma^2$ is the estimated variance due to instrument noise. FP$_{\text{mean}}$ is the proportion of false positives obtained for this bead population if we threshold at the mean, computed using equation \eqref{eqn:FPmean}.} 
\begin{tabular}{ c c| c c c }
 Bead size & Flow rate & rel. noise ($\frac{\sigma^2}{\nu^2}$) & rel. error ($\frac{\sigma}{\mu}$) & FP$_{\text{mean}}$\\
\hline
 $7 \mu\text{m}$ & 22$\mu$L/min & 0.17 & 0.024 &  0.135 \\ 
 $15 \mu\text{m}$  & 29.5$\mu$L/min & 0.05  & 0.018 & 0.074 \\   
\end{tabular}
\label{table:sorter}
\end{table}

Knowledge of the noise variance lets us estimate the proportion of the sorted population in our experiment whose true fluorescence value actually lies above the gate. Viewed as a classification problem, this is the false positive rate. 
Using our notation from the previous section,the false positive rate $\text{FP}(t)$ with gate $t$ for a normally distributed population is given by
\begin{equation}
\text{FP}(t) = P( X >t | X+ \epsilon <t) = 1 - \frac{\Phi_2\left( \frac{t-\mu}{\lambda} ,\frac{t-\mu}{\nu}, \frac{\lambda}{\nu} \right)}{\Phi\left(\frac{t-\mu}{\nu}\right)}.
\end{equation}
In particular, we compute the false positive rate when the gating threshold is equal to the measured mean, a quantity we call the \textit{the false positive rate at the mean} ($\text{FP}_\text{mean}$), as a proxy for the reliability of the measured values.
\begin{equation}\label{eqn:FPmean}
\text{FP}_{\text{mean}}=  1 - 2\Phi_2\left( 0,0,\frac{\lambda}{\nu}\right)
\end{equation}

\begin{figure}[h]
\begin{minipage}[b]{0.5\linewidth}    \includegraphics[width=\linewidth]{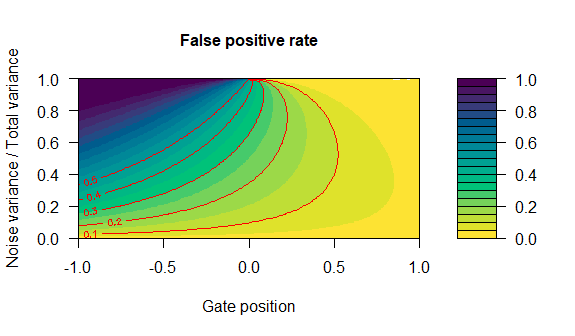}
\end{minipage}
\begin{minipage}[b]{0.5\linewidth}
\includegraphics[width=\linewidth]{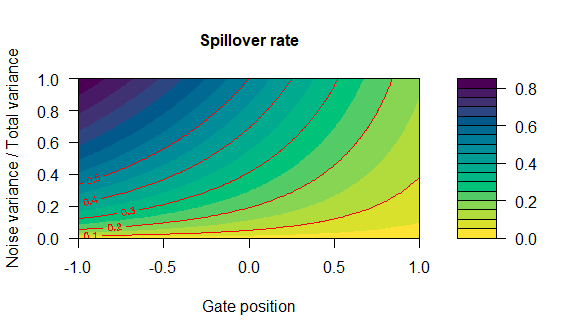}   
\end{minipage}
  \caption{Contour plot of the false positive and spillover rates for a normally distributed population as a function of the gate position and the relative noise variance. The gate position is expressed in terms of standard deviations from the mean.}	
    \label{fig:spillfp}
\end{figure}

Note that the false positive rate described above arises entirely due to errors in the measured fluorescence values, and is not a measure of the purity or yield of the sorter itself.
Although related, the false positive rate is not the same thing as the fraction of remeasured beads which fall above the gating threshold (the \textit{spillover ratio}), as illustrated by the contour plots in Figure \ref{fig:spillfp}. Both quantities tend to $0$ and $1$ at $+\infty$ and $-\infty$ respectively, as a natural consequence of the fact that we are capturing all (or none) of the population. It is perhaps tempting to directly use the spillover rate as a proxy for the noise variance, but as Figure \ref{fig:spillfp} shows, the relative noise variance is substantially higher than the spillover rate. Additionally, it is important to note that the false positive rate as computed here uses the fact that the underlying population has a Gaussian distribution. 

\subsection{Noise estimates for other cytometers}

\begin{figure}[ht]
\centering
    \includegraphics[width=\linewidth]{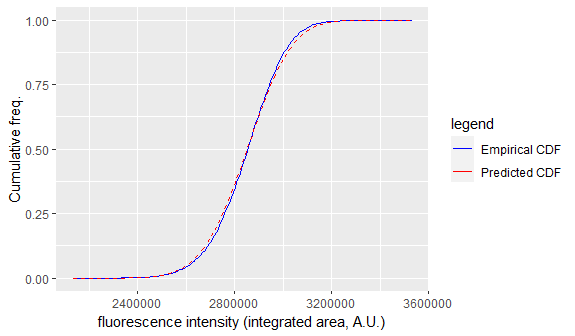}
     \caption{Predicted vs empirical CDF of the post sort distribution for the $7$ micron beads on Instrument 2, showing a maximum difference of $~3\%$.}	
    \label{fig:cdf_Instrument 2}
\end{figure}

We remeasured the pre and post sort beads on different cytometers, which we will label Instrument 2 and 3. The $7$ micron beads were measured on Instrument 2, while the $15$ micron beads were measured on both Instrument 2 and 3. Using the previously obtained noise estimates from the sorter, we use equations \eqref{eqn:Aest},\eqref{eqn:Best},\eqref{eqn:sigma2} to obtain estimates of the scale coefficients $A$ and $B$ and the noise inherent to the instruments, which we summarise in Table \ref{table:Instrument 2}. The substantial variation in the scale coefficients strongly indicates that our affine transformation hypothesis is only local in nature, and does not hold across the full range of intensity values. See section \ref{disc} for a more detailed discussion of this concept, including practical implications. 
Using the expression in equation \eqref{eqn:KII} for the post-sort cumulative densities, we partially validate our analysis by comparing the predicted and empirical cdfs. Figure \ref{fig:cdf_Instrument 2} plots these for one trial of the $7$ micron beads run on Instrument 2.

\begin{table}[h]
\centering
\caption{Estimated scale coefficients and measures of uncertainty for different bead sizes on Instrument 2. Respectively, $\sigma^2,\nu^2,\mu$ are estimated noise variance, measured pre-sort variance, and measured pre-sort mean. FP$_{\text{mean}}$ is computed using equation \eqref{eqn:FPmean}.} 
\begin{tabular}[h]{ c c |  c c c c c }
 Bead size & Flow rate & A & B  & rel. noise $\frac{\sigma^2}{\nu^2}$ & rel. error $\frac{\sigma}{\mu}$ &  FP$_{\text{mean}}$ \\
\hline
 $7 \mu\text{m}$ &  90$\mu$L/min & 16.6  &  773883 & 0.87  & 0.09 & 0.29\\ 
 $15 \mu\text{m}$  & 120$\mu$L/min & 54.9 & -1665643 &  0.68 & 0.15 & 0.30 \\   
\end{tabular}
\label{table:Instrument 2}
\end{table}

\begin{table}[h]
\centering
\caption{Estimated scale coefficients and measures of uncertainty for the 15 micron bead on Instrument 3. } 
\begin{tabular}[h]{ c| c c c c c  }
  Flow rate & A & B  & rel. noise $\frac{\sigma^2}{\nu^2}$ & rel. error $\frac{\sigma}{\mu}$ & FP$_{\text{mean}}$   \\
\hline
  4.02$\mu$L/min & 0.0032 & -92.9 & 0.019 & 0.014 & 0.044 \\   
\end{tabular}
\label{table:crossinst}
\end{table}


\subsection{Noise as a function of flow rates}
\begin{figure}[ht]
\includegraphics[width=\linewidth]{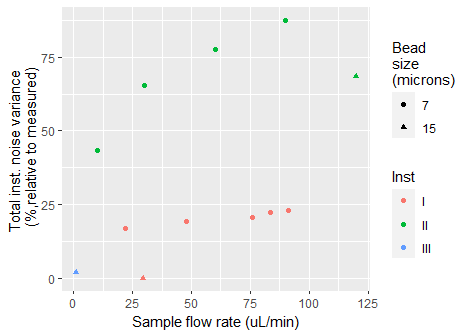}
\caption{Total instrument-induced noise variance (expressed as a percentage of the measured variance), for 7 and 15 micron beads measured on three instruments (I: sorter, II: a commercial analytic cytometer, III: a custom `serial' cytometer, described in \cite{disalvo2022serial}). Sample flow rates for the sorter and Instrument III are experimentally derived estimates.}
\label{fig:flowrates}
\end{figure}

In a flow cytometer, the \textit{flow rate} of the sheath fluid substantially affects the focusing of the particles \cite{disalvo2022serial}, which in turn affects the measured intensities. The flow rate of the sample is typically an order of magnitude smaller than the sheath fluid flow rate. Nonetheless, variation in the sample flow rate does lead to observable differences in instrument noise, as we show using our framework. Figure \ref{fig:flowrates}  depicts the total instrument-induced noise variance for 2 sets of beads on 3 different instruments, expressed as a function of sample flow rate. 

\section{Discussion}\label{disc}
As discussed earlier, if it were possible to obtain repeated measurements of a particular particle, the problem of estimating the noise would be much easier (assuming the noise is homoscedastic): one simply takes the variance of the repeated individual measurements. Such measurements are not possible in commonly used cytometers, but an approximation can be obtained using the recently developed serial cytometer \cite{disalvo2022serial}, which utilizes particle tracking to take repeated measurements of an individual particle over multiple laser interrogation regions.  A key feature of our analysis is the fact that we only require repeated measurements \textit{in the aggregate}. These noise estimates have a number of practical applications, which we discuss below. We end with an overview of the assumptions, limitations and possible extensions of our model.

\subsection{Applications}

\subsubsection*{Noise estimates as a measure of instrument or bead quality}
The noise estimates we obtain are, in some sense, a measure of instrument quality. However, in order to meaningfully compare two instruments, one needs to obtain noise estimates using identical sets of beads measured under constant conditions. (See Tables \ref{table:sorter},\ref{table:Instrument 2} and Figure \ref{fig:flowrates}, where we obtain noise estimates for different bead types and instruments, albeit under varying flow rates.)  
On the other hand, our analysis also allows us to compute the variance solely due to population level effects. High population variance in this case is indicative of high variance in the number of active fluorophores per bead. Our methods therefore allow us to compare \textit{bead quality}, by analyzing the relative signal variances of different sets of beads of the same nominal diameter and mean fluorescence intensity. This is not novel - cytometrists often use the CV of reference beads or cells to assess their quality (\cite{wang2021establishing}). Our analysis allows us to refine this idea, by disentangling the contribution of the instrument and the population towards the measured CV.
As a note of caution, we warn that all estimates are subject to model-form errors, and any measure of quality obtained by this process should be cross-validated by other methods.

\subsubsection*{Identifying sources of noise}
Our models only estimate the total noise present in the instrument, which is likely a sum of noise contributed by multiple sources. As we show with our analysis of flow rates (Figure \ref{fig:flowrates}), one may estimate the relative effect of a certain parameter by repeating our sorting experiment over a range of parameter values, all else being constant. There are a number of other variables which potentially affect the noise, such as particle shape, photodetector type, gain setting, or the density of the particles in the sheath fluid. Potential avenues for future work include a multi-component correlation study, to disentangle these effects from other sources of noise present in an instrument.

\subsubsection*{Gating and classification errors}
\begin{figure}
    \begin{minipage}[b]{0.48\linewidth}
            \includegraphics[width=\linewidth]{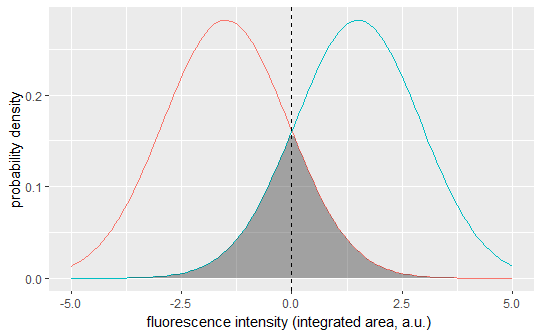}
            
    \end{minipage}
    \begin{minipage}[b]{0.48\linewidth}
            \includegraphics[width=\linewidth]{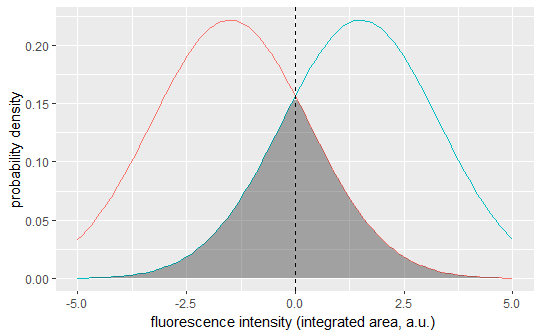}
    \end{minipage}
    \caption{An example showing the decrease in classification accuracy caused by increased noise. (Left) Intensity distributions for two normally distributed populations of means $-1.5,1.5$ and variance $1$, with instrument noise of variance $1$. (Right) Intensity distributions for the same populations, but with instrument noise variance $1.5$. The minimal theoretically possible classification error (the Bayes error \cite{hand2001idiot}) corresponds to the shaded areas under the curves, and increases from $15\%$ on the left to $21\%$ on the right.   }    
    \label{fig:classaccex}
\end{figure}

One can also use noise estimates to better understand gating or classification errors.
Let us illustrate this with the help of an example. Assume that there are two populations of interest whose members one wishes to distinguish. Assume a reference laboratory measures these populations and characterizes their fluorescence intensity distributions, which is later used to classify new samples. This classification might conceivably be performed under different instrument settings, such as a higher sample flow rate. As shown in Figure \ref{fig:flowrates}, sample flow rates have significant effects on instrument noise. It should be clear that optimal classification error rates increase with instrument noise (see Figures \ref{fig:classaccex}). Consequently, we can now quantify the decrease in optimal classification accuracy due to increased sample flow rates. 

\subsection{Limitations and possible extensions}

\subsubsection*{Local versus global effects}
We have relied on a number of assumptions in this work about the nature of the instrument noise. A significant caveat is that these assumptions are only \textit{local} in nature (cf.\cite{wood1998evaluating}) - i.e. they do not hold over the full range of the instrument. To begin with, we assumed that the noise and the signal are independent random variables. For our analysis to work, this needs to be true only over a small interval corresponding to the range of values obtained from one bead.  Table \ref{table:sorterraw} shows that the estimated noise variance does vary over our different domains. A full accounting of this effect might need to involve a model which incorporates some dependence of the noise on the magnitude of the signal.

\subsubsection*{Gain-dependent effects and axis-scaling between instruments}

The affine relationship that we posited above is also a local phenomenon, as is seen from the fact that the scale coefficients in Table \ref{table:Instrument 2} show substantial variation. In general, we expect that the coefficients are a function of the measurement value. In order to obtain a  comprehensive noise profile for a cytometer using our model, one ought to repeat the sorter-based experiment for beads of varying brightness under different instrument parameters, such as the gain. This will yield a dataset of instrument noise as a function of intensity and other parameters, which may then be analysed using tools from multivariate statistics to obtain a functional relationship between these quantities. In future work, we aim to quantify the gain-dependence of instrument noise, and develop more sophisticated models for axis-scaling between instruments.

\subsubsection*{Uncertainties in the sorting process}

An implicit assumption in our work is that the sorter works perfectly - i.e. the \textit{purity} of the sorted sample is very close to $100\%$. This is a fairly reasonable assumption in practice for most commercial sorters (cf \cite{bdtech}). However, our numerical experiments suggest that a $2\%$ error in the gating threshold $\gamma$ leads to an error of the order  of $5\%$ in the estimated noise variance.  More sophisticated models might therefore incorporate a probabilistic gating threshold, which accounts for the fact that a certain fraction of particles are incorrectly sorted. Note that we do not make any assumptions on the \textit{yield} - even if the sorter only captures a fraction of the particles with values below the threshold, the model still works, as this just corresponds to a subsample of the actual distribution.

\subsubsection*{Non-Gaussian signals and noise}

The estimator in \eqref{eqn:SDEstimator} is formulated for Gaussian signals and noise. Our experiments and analysis indicate that this is justified, at least in the case of tightly clustered beads of uniform size. However, measured fluorescence intensities of populations consisting of cells or beads of varying diameter do not necessarily obey a Gaussian distribution. As we outline in the appendix, it is possible to obtain a noise estimate even in these cases, provided we assume that the noise in question is Gaussian. 
It is naturally tempting to ask if one can recover the `true' measurement signal using our estimates. This is a deconvolution problem, which is mathematically ill-posed \cite{fan1991optimal,matias2002semiparametric,masry1992gaussian}), and is vulnerable to model-form errors. A more practical approach is to use these noise estimates to better understand the limitations of gating or classification strategies on specific instruments.

\section*{Data Availability}
Data and code developed as part of this work are available upon reasonable request.

\section*{Declarations of Competing Interests}
The authors have no competing interests to declare.

\section*{Disclaimer}
Certain commercial equipment, instruments or materials are identified in this paper to foster understanding. Such identification does not imply recommendation or endorsement by the National Institute of Standards and Technology, nor does it imply that the materials or equipment identified are necessarily the best available for the purpose.

\bibliographystyle{unsrt}
\bibliography{ref.bib}

\appendix
\appendixpage
\section{Estimators for non-Gaussian signals}

Let $f,g,h$ denote the probability density functions of $Y_1,X,\epsilon_1$ respectively, and let $\gamma$ denote the gating threshold. We seek to obtain the probability density function $k$ corresponding to the second set of measurements in our experiment - i.e. we want
\[ k(x) = \frac{d}{dx} P(Y_2 <x | Y_1 < \gamma).\]
Conditioning on the value of $Y_1$, we get 
\begin{eqnarray}
    P(Y_2 < x | Y_1 < \gamma ) = \int_{-\infty}^{\gamma} P(Y_2< x | Y_1 = y) P(Y_1 = y | Y_1 <\gamma) dy 
\end{eqnarray}
where 
\begin{equation}
    P(Y_1=y|Y_1< \gamma) = \left(\int_{-\infty}^{\gamma} f(s) ds\right)^{-1} f(y) := C^{-1} f(y).
\end{equation}
Conditioning on the value of $X$, we get 
\begin{equation}
P(Y_2 < x | Y_1 =y) = \int_\mathbb{R} P(Y_2 < x | Y_1=y, X=r) P(X=r | Y_1 = y) dr.    
\end{equation}
Since $X,\epsilon_1$ are independent,
\begin{equation}
    P(X=r | Y_1 = y) = \frac{P(X =r , Y_1 = y)}{f(y)} = \frac{g(r)h(y-r)}{f(y)}.
\end{equation}
\begin{equation}
    P(Y_2 < x | Y_1 = y, X=r) = \frac{P(\epsilon_2<x-r,X=r,\epsilon_1=y-r)}{g(r)h(y-r)} = \int_{-\infty}^{x-r} h(s) ds.
\end{equation}
Putting it all together, we get
\begin{align}\label{eqn:K}
    K(x)= P(Y_2<x | Y_1 <\gamma)= C^{-1} \int_{-\infty}^{\gamma} dy \int_\mathbb{R} dr \int_{-\infty}^{x-r} ds \,  h(s)  g(r) h(y-r) 
\end{align}
We may take the derivative in order to obtain the probability density function $k$:
\begin{equation}
k(x) = C^{-1}\int_{-\infty}^{\gamma} \int_\mathbb{R} h(x-r)h(y-r) g(r) dr\, dy.
\end{equation}
Letting $H(r)$ denote the cumulative density of $\epsilon$, this becomes 
\begin{equation}\label{eqn:k}
    k(x) = C^{-1} \int_\mathbb{R} h(x-r) g(r) H(\gamma-r) dr =(h \star \alpha)(x),
\end{equation}
where 
\begin{equation}\label{eqn:alpha}
    \alpha(x)=C^{-1} g(x)H(-x), \; \; C=\int_{-\infty}^{\gamma} f(s) ds
\end{equation}

For what follows, we assume that the pdf of the noise, $h$, is more sharply peaked when compared to $g$. Using the fact that $f = g \star h$, we approximate $g$ as a second order Taylor polynomial to obtain the following `Laplace style' approximation:
\begin{align}
    f(x) & = \int_{\mathbb{R}} h(t) g(x-t) dt \sim \int_{\mathbb{R}} h(t) \left( g(x) - g'(x) t + g''(x) \frac{t^2}{2} \right) dt  \nonumber \\
    & = g(x) + g''(x) \frac{\sigma^2}{2} \label{eqn:fapprox}
\end{align}

Note that the ODE given by ~\eqref{eqn:fapprox} does not come with a initial or boundary condition - instead, we require that the solution is a probability density function, so $g(x)\geq 0 $ and $\int g =1$. Although it is possible to solve the ODE exactly, it is difficult to determine the coefficients of the general solution which ensure that $g$ is positive. It turns out that a perturbative approach is more suitable for our purposes here. If we assume that the solution $g$ has the form

\[
g(x) = \sum_{n \geq 0} g_n(x) \sigma^n,
\]
and solve for $g_n$, we obtain
\[
g_0(x) = f(x), \; g_1(x)=0, \; g_2(x) = -\frac{f''(x)}{2}.
\]
We then have the following approximation for $g$:
\begin{equation}\label{eqn:g_approx}
g(x) \sim f(x) - f''(x) \frac{\sigma^2}2.
\end{equation}
Note that a similar approach for $k$ is not advisable, as $\alpha$ rapidly varies around the cutoff point $0$. We get around this by first approximating $H$ by a polynomial. Let 
\[
p(x) = \sum_{i=0}^n c_i x^i
\]
be a polynomial approximation of the standard normal cdf with some specified error $\delta$ on an interval $[-T,T]$. Then 
\begin{equation}\label{eqn:polycdf}
p_\sigma(x) = \sum_{i=0}^n \frac{c_i}{\sigma^i} x^i
\end{equation}
is the corresponding approximation for $H$ (the normal cdf with variance $\sigma^2$) on the interval $[-T\sigma, T\sigma]$.

Using the polynomial approximation for $H$ and the second order Taylor approximation for $g$, we get

\begin{align}
C k(x) & \sim \int_{-T\sigma}^{T\sigma} h(t) \left( g(x) - g'(x) t + g''(x) \frac{t^2}{2}\right)\left(\sum_{i=0}^n \frac{c_i}{\sigma^i} (t-x)^i\right) dt \\
& \sim \sum_{i=0}^n \sum_{j=0}^i (-1)^{i-j} \frac{c_i}{\sigma^i} {\binom{i} {j}} x^{i-j} \int  h(t) \left( g(x) - g'(x) t + g''(x) \frac{t^2}{2}\right) dt \nonumber \\
& = g(x) A_0\left(\frac{x}{\sigma}\right) + g'(x) A_1\left(\frac{x}{\sigma}\right)\sigma + g''(x) A_2\left(\frac{x}{\sigma}\right)\frac{\sigma^2}{2}. \label{eqn:kapprox}
\end{align}

where $A_0,A_1,A_2$ are polynomials whose coefficients are functions of $c_i$. This follows from the fact that the $n$th moment of $h$ is a constant multiple of $\sigma^n$. Explicitly, we have

\begin{align}
A_0\left(\frac{x}{\sigma}\right) & = \sum_{i=0}^n \sum_{\substack{j=0 \\ \textrm{j even}}}^i (-1)^{i-j} (j-1)!! {\binom{i} {j}} c_i  \left(\frac{x}{\sigma}\right)^{i-j} \label{eqn:A0} \\
A_1\left(\frac{x}{\sigma}\right) & =  \sum_{i=0}^n \sum_{\substack{j=0 \\ \textrm{j odd}}}^i (-1)^{i-j+1} j!! {\binom{i} {j}} c_i  \left(\frac{x}{\sigma}\right)^{i-j} \label{eqn:A1} \\
A_2\left(\frac{x}{\sigma}\right) & = \sum_{i=0}^n \sum_{\substack{j=0 \\ \textrm{j even}}}^i (-1)^{i-j} (j+1)!! {\binom{i}{j}} c_i  \left(\frac{x}{\sigma}\right)^{i-j}
\end{align}

$\eqref{eqn:kapprox}-A_2\left(\frac{x}{\sigma}\right) \eqref{eqn:fapprox}$ gives 

\begin{equation}
Ck(x) - A_2\left(\frac{x}{\sigma}\right) f(x) \sim g(x)\left(A_0\left(\frac{x}{\sigma}\right)-A_2\left(\frac{x}{\sigma}\right)\right) + g'(x)A_1\left(\frac{x}{\sigma}\right) \sigma
\end{equation}
Utilizing the approximate solution of $g$ we obtained in ~\eqref{eqn:g_approx}, we get

\begin{align}
L_x:= Ck(x) - A_2\left(\frac{x}{\sigma}\right) f(x) -  & \left(f(x) - f''(x) \frac{\sigma^2}{2}\right)\left(A_0\left(\frac{x}{\sigma}\right) -A_2\left(\frac{x}{\sigma}\right)\right) \nonumber \\- &\left(f'(x) - f'''(x) \frac{\sigma^2}{2}\right)A_1\left(\frac{x}{\sigma}\right) \sigma = 0.\label{eq:objfn}
\end{align}

The values of $k(x)$ and $f(x)$ and its derivatives can be estimated by using kernel density estimates. Consequently, for each $x_0$, we obtain a rational objective function $L_{x_0}^2$ which we minimize to get an estimator $\widehat{\sigma}$ for $\sigma$. Assuming that the ratio of noise to signal standard deviations is no greater than $1$, we obtain the relative noise standard deviation by scaling our dataset and minimizing the objective function over the domain $(0,1]$.

\begin{equation}
\widehat{\sigma}:= \min_{s \in (0,1]} L_{x_0}^2(s).
\end{equation}
\end{document}